# Carbon nanotube based biomedical agents for heating, temperature sensoring and drug delivery

Rüdiger Klingeler (r.klingeler@ifw-dresden.de)

Silke Hampel

Bernd Büchner

Leibniz Institute for Solid State and Materials Research (IFW) Dresden, Helmholtzstr. 20, D-01069 Dresden, Germany

## Abstract

Due to their extraordinary physical and chemical properties carbon nanotubes reveal a promising potential as biomedical agents for heating, temperature sensoring and drug delivery on the cellular level. Filling carbon nanotubes with tailored materials realises nanoscaled containers in which the active content is encapsulated by a protecting carbon shell. We describe different synthesis routes and show the structural and magnetic properties of carbon nanotubes. In particular, the filling with magnetic materials offers the potential for hyperthermia applications while the insertion of NMR active substances allows the usage as markers and sensors. The potential of carbon nanotubes for biomedical applications is highlighted by hyperthermia studies which prove their applicability for local in-situ heating. In addition we have shown that a non-invasive temperature control by virtue of a carbon-wrapped nanoscaled thermometer and filling with anti-cancer drugs is possible.

Strong adverse effects on the healthy tissue in the vicinity of a tumour are a major drawback in current cancer therapies. One innovative technological approach to solve this problem focuses on therapies on the cellular level by applying intracellular probes, i.e. the transfer of nano-sized biocompatible devices into the cells. Most promising materials in this regard are magnetic nanoparticles which can be addressed by external magnetic fields. A great advantage of magnetic fields is their biocompatibility, i.e. they penetrate tissue non-invasively and without known adverse effects, and their weak interaction with organic matter so that deep layers of (human) tissue can be reached. In particular, magnetic nanoparticles can be localized in deep tissue, external static magnetic fields can fix them at a precise position, gradient fields can move them and alternating (AC) fields lead to local heating. The latter can be utilized for hyperthermia applications which is shown e.g. by the use of superparamagnetic nanoparticles for magnetic fluid hyperthermia [1] Much of the current research is focused on iron oxide nanoparticles which have proven their feasibility in animal experiments [2,3] and are now under clinical trials [4]. Other magnets usually present limitations due to biocompatibility issues, since most of them contain potentially toxic elements.

Due to its particular magnetic properties, i.e. large anisotropy and saturation magnetisation, metallic iron could generate heat more efficiently in comparison to iron oxides since various mechanisms yield dissipative effects such as domain wall motion etc. [5]. In practice, higher heating efficiency means that less nanoscaled material would have to be introduced into the biological system in order to achieve the targeted hyperthermia effect. The use of nanoparticles made of iron, however, is hindered by the fact that oxidation in ambient or biological conditions has to be avoided. A promising way to overcome this problem appears to be the coating of the iron with a carbon shell by insertion of the material inside of carbon nanotubes (CNT) and thereby protecting the biological environment and the filling material against each other. Degradation of filling materials is avoided and their potential toxicity and adverse effects are suppressed so that CNT provide a smart carrier system on the nanometer scale.

A targeted hyperthermia therapy using iron filled CNT and, in general, utilisation of tailored carbon nanotube based biomedical agents for therapeutic and diagnostic purposes can hence be envisioned which will be detailed in the following sections. After introducing some basic properties of CNT, a short review of biofunctionalisation, biocompatibility and toxicity aspects is provided. The synthesis of iron filled CNT as well as magnetic and hyperthermia studies are discussed. Adding additional material such as a thermometer in Fe-filled CNT increases the potential of CNT as hyperthermia inducing agents since simultaneous local heating and temperature control might be feasible. Moreover, the container feature of CNT offers the step beyond the controlled heat treatment but also suggests the additional local release of therapeutics.

## 2. Biocompatibility and Toxicity

CNT are hollow carbon structures with one or more walls, a small diameter on the nanometre scale and a large length in comparison. Their mechanically and chemically stable carbon shells can be opened, filled and closed again without losing their stability. Experiences in filling CNT range back to their discovery in 1991 [6]. Since then, extensive work has been performed to synthesise CNT and to functionalise them both exohedrally, i.e. by attaching functional elements to the outer shell, and endohedrally by filling with various materials. CNT can be filled with metals, semiconductors, salts, organic materials, fullerenes, etc., either during the synthesis process or through subsequent opening, filling, and closing of the CNT. In particular, ferromagnetic materials such as Fe, Ni, and Co can be encapsulated which is relevant to hyperthermia applications. Introducing fluorescent markers offers a route for visualising CNT [7], which is also possible by exohedral functionalisation [8,9]. The container feature of CNT allows, in principle, simultaneous filling of CNT with different materials thereby combining multiple functionalities in one kind of carrier (Fig. 1). In this way CNT provide a smart carrier system on the nanometer scale which can be filled with tailored materials to address specific purposes. The container function is underlined by recent model calculations for encapsulation of the anticancer drug cisplatin. Interestingly, all three orientations of cisplatin might be transferred into a CNT if its minimum radius is at least 4.785 Å while the maximum uptake occurs for radii of approximately 5.3 Å. These results hence suggest that multi-walled CNT with a relatively large inner diameter are particularly appropriate for drug delivery of common therapeutics like cisplatin or carboplatin [10].

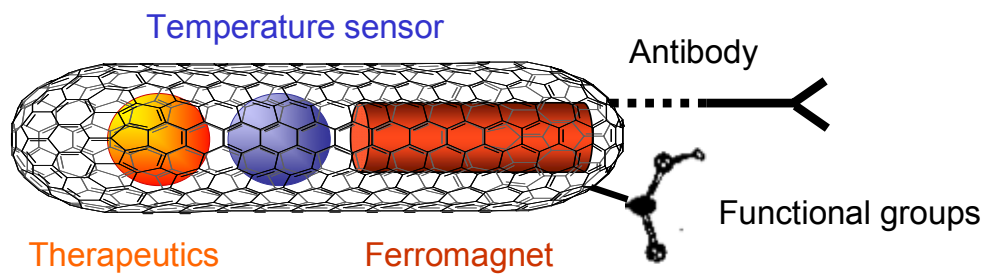

**Fig. 1:** Sketch of a filled carbon nanotube serving as multi-functional container for in-vivo applications. A ferromagnet can induce heat in AC magnetic fields and a material with strongly temperature dependent nuclear magnetic resonance (NMR) signal might serve as a thermometer. Additional drug delivery can be envisaged. Exohedral functionalisation to achieve biocompatibility is sketched. Note, that single-walled as well as multi-walled CNT can be realised.

Beyond the shielding effect against a biological environment, the carbon coating offers an interface for further exohedral functionalisation with suitable (bio-) molecules. A major task of such functionalisation is the stable dispersion of the nanoparticles in aqueous media which still is a major challenge. Both non-covalent and covalent modification strategies can be applied among which the former preserve the pristine CNT structure while covalent modification introduces partial damage of the outer wall but in general yields better dispersion. Dispersion becomes even more crucial if ferromagnetically filled CNT are envisaged which exhibit an increased tendency to agglomerate due to magnetic interactions. In addition, exohedral functionalisation is needed to get the highly symmetric carbon structures compatible to actual biological environments. Nowadays, also functionalisation aiming to perform dynamic tasks such as target recognition, target transformation, transport, or electrical conduction in the living cell is addressed. These functions can be provided by biomolecules, like DNA and enzyme proteins. Importantly for any therapeutical or biomedical usage, functionalised CNT can cross effectively biological barriers such as the cell membrane and penetrate the individual cell. Recently, it was shown that DNA wrapped SWNT were enveloped by cancer cells and they were used to deliver a lethal dose of microwave radiation to the cancer cells [11]. In contrast to such a wrapping CNT with biomolecules, any chemical exohedral functionalisation needed to improve the biocompatibility of CNT will structurally perturbate the external wall which however in the case of multi-walled CNT does not affect the overall carbon shielding. A variety of methods exist for the exohedral functionalization of carbon nanotubes [12,13], some of which have been successfully applied for the conjugation of proteins, drugs and fluorescent dyes [14,15,16,17] and active tumor targeting *in-vivo* [18]. Successful biofunctionalisation is e.g. demonstrated by Pantarotto et al. who observed that functionalized CNT can cross the cell membrane and accumulate in the cytoplasm or reach the nucleus without being toxic for the cell [15]. Soluble CNT can be coupled with amino acids and bioactive peptides [19] for further derivatization. It was also shown that stable complex formation of CNT and cationic lipid may accelerate the delivery of nanotubes into the bladder cancer human cells [20]. For the actual mechanism of internalization, i.e. endocytosis or phagocytosis, which has been discussed contradictorily, the actual morphology of the CNT seems to be crucial [14,15,21].

Despite various industrial applications of CNT and their large scale synthesis only little is known about the interaction of CNT with biological environments. CNT and other nanomaterials when placed in contact with human (or other mammalian) body fluids and tissues are recognised by the immune system complement proteins, and may also interact with other systems, such as coagulation proteins and cell-surface proteins. Binding of complement proteins activates the complement system via both classical and alternative pathways and results in strong binding of several complement proteins to the CNT [22]. Such complement activation influences subsequent interaction of the CNT with cells and tissues and is a predictor of potential toxicity in animal models. Salvador-Morales et al. report that such protein binding to CNT is highly selective. In particular, fibrinogen and apolipoproteins were the proteins that bound to CNT in greatest quantity. Among proteins contained in lung surfactant, SP-A and SP-D selectively bind to CNT so that chronic level exposure may result in sequestration these proteins [23].

Although a variety of toxicity studies has been published (for recent reviews see e.g. [24,25]), no clear picture evolves for CNT in general. This is e.g. illustrated by contrasting *in-vitro* studies showing CNT to be safe [26,27] or to induce significant toxic effects [28,29,30]. One reason for this ambiguity is the fact that – similar to other nanoparticles – a large number of specific factors govern the toxicity of CNT such as their shape and size (diameter, length), the number of shells (single- or multi-walled CNT), agglomeration state and surface chemistry [31]. In particular, the concentration of CNT is not necessarily a main parameter. On the other hand, ambiguity also results from a lack of standardisation and of thorough characterisation of, e.g., defects of the outer shell or potentially toxic contaminants. Such non-carbon material originating from the synthesis process of pristine empty CNT might amount to 5-10% of the total mass. This presumably accounts for much of the reported differences so that details of synthesis, choice of catalyst particles, washing procedures and dispersion methods have to be considered thoroughly.

In the following we hence concentrate on multi-walled Fe-filled CNT which synthesis and properties are described in §3 and §4. We recall the feasibility of an efficient delivery of Fe-filled CNT into human cancer cell lines which has been shown in cell culture experiments [20]. A pre-treatment of Fe-filled CNT with cationic lipid was found to cause a qualitative delivery of the complexes into the cytoplasm but not into the nucleus. A recent study on cytotoxic effects of Fe-filled CNT *in-vitro* addressed metabolic activity, cell proliferation, apoptosis and cell cycle distribution of a malignant (PC-3) and a non malignant (fibroblasts) cell lineage. The data imply that CNT strongly associate with cells within a short incubation period. The presence of CNT in cells did not pose any significant toxic effect [32]. This observation is corroborated by an animal study on mice which indicated no remarkable toxic or adverse effects over a period of 440 days after the intraperitoneal or intravenous injection of pristine Fe-filled CNT. In particular, based on TEM and histological analyses after 6 weeks Mönch et al. [33] report the absence of any indication on inflammation. In one case, CNT have been administered up to a total of >1g Fe-filled MWCNTs/kg body weight over a period of 3 months. Interestingly, in agreement with the fact that pristine CNT were used, large agglomerates have been observed in various organs in case of the intravenous treatment. However, such agglomerates seem to have no drastic effects. All animals survived for more than one year and showed no abnormalities in behaviour or weight suggesting a general biocompatibility of the CNT for the applied doses.

The long-term absence of significant toxicity effects in the mentioned animal study also implies the expected insignificant degradation of CNT in the biological environment. Regarding their fate in-vivo, excretion in feces and urine has been reported after injection of empty CNT injected intravenously or intraperitonally in mice [34,35]. Studies on the biodistribution and translocation pathways of empty CNT in mice, however, indicate accumulation predominately in liver and retain for long time while low acute toxicity is confirmed [36].

## 3. Synthesis of Ferromagnetic Filled Multi-walled Carbon Nanotubes

Encapsulation of iron nanowires in CNT realises highly anisotropic ferromagnetic nanoparticles which are discussed for a wide range of potential applications [37]. Among a variety of methods to synthesize these filled carbon structures such as arc discharge [38] and laser ablation [39], the chemical vapour deposition (CVD) is applied when a high yield uniform multi-walled CNT is aimed [40].

A suitable method for synthesizing CNT filled with metals as Fe, Co or Ni is the so-called ''in situ method'', a special CVD-method, in which the formation of CNT and their filling with ferromagnetic elements or compounds take place simultaneously. Here, relevant precursors are needed, which e.g. provide as well iron for the filling as carbon for the shells. The most common precursors are metallocenes $[Me(C_5H_5)_2;$ Me = Fe, Co, Ni] which supply a large scale of well-defined multi-walled (MW) CNT with a high filling yield (about 50 %) of the ferromagnetic material. Many groups world-wide use this method in different equipments. Sen et al. reported the first experimental results for the synthesis of ferromagnetic MWCNT [41]. In a typical experiment the ferromagnetically filled MWCNT are grown in a quartz tube reactor inside a dual zone furnace system. The main parameters of the deposition process are the sublimation temperature of the metallocene in the first furnace, the gas flow rate, and the temperature of the second hot furnace zone where the pyrolysis of metallocene and the deposition of filled tubes take place. CNT are formed on oxidized silicon substrates uncoated or coated (2 nm layer of different materials such as Fe, Co and Permalloy) placed inside the reaction zone. In this case they grow in perpendicular alignment to the substrate surface. Much more nanotubes are formed on the wall of the quartz tube reactor (which is a mixture of bundles of well-aligned Fe-filled MWCNT) [42,43, 44, 45].

An additional technique is the liquid source CVD (LS-CVD) [40]. This method is characterized by a constant and reproducible transport of the precursor. The metallocene, especially the ferrocene, is dissolved in cyclopentane and dropped on the moving band continuously. In the first part of the system the solvent is vaporized and only the ferrocene is transferred into the reactor at a defined temperature and with a constant transport velocity. In the deposition reactor a second moving band, populated with precoated substrates is positioned. This LSCVD method is highly suited for the continuous production of defined ferromagnetically filled CNT. Wang et al. have synthesized Fe-filled CNT with a high filling ratio by using ferrocene and dichlorbenzene as precursor [46]. This solution is injected through a nozzle directly into the reactor so that a spontaneous decomposition reaction occurs. Pichot et al. applied an aerosol assisted CVD (AA-CVD) [47] device to produce Fe filled CNT. Ultrasonicating a solution of ferrocene in cyclohexane produced an aerosol, which was then carried for 15 min by an argon flow through a quartz reactor placed in a tubular furnace at 850 °C. Eventually, MWCNT carpets were deposited on the reactor walls [48].

Filled CNT can be also synthesised by a post-synthesis method. This technique includes (1) the synthesis of empty CNT, (2) the opening of their ends, (3) a filling step with various materials (metals, salts, therapeutics) and (4) the closing step of the filled CNT. The opening procedure is mostly done by well established wet-chemistry techniques [49] or by oxidation in air [50, 51]. The easiest way to incorporate the material into the open ended CNT is over a gasphase reaction where the CNT and the filling material are inserted together into glass ampoules sealed under vacuum and heated beyond the sublimation temperature of the filling material [52, 53]. But also a wet chemical approach can be applied where capillarity is the driving force [54,55],49. The closure of the openended and filled CNT can be realised by a redeposition with a polymer or other carbon-containing phases [56]. Various techniques such as transmission electron microscopy and X-ray diffraction demonstrate the successful synthesis of Fe-filled CNT by our CVD technique [40,43,42,57,58,59,60]. E.g., the cross section shown in Fig. 2 clearly indicated the single crystalline Fe filling as well as the carbon shell.

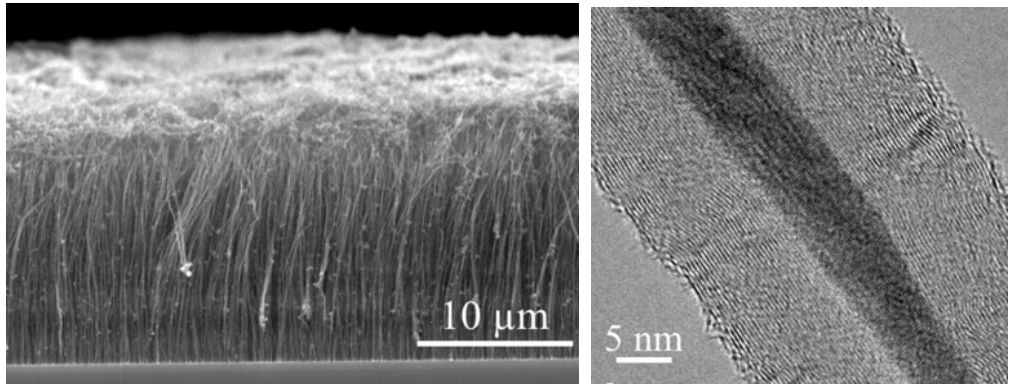

**Fig. 2:** Typical SEM and TEM images of well-aligned iron filled multi-walled carbon nanotubes synthesised by LS-CV.

The length of CNT is in the micrometer range and outer diameters vary from 20-80 nm. The Fe-filling is discontinuous and consists of iron nanowires with a filling yield of ~50%. Interestingly, by x-ray diffraction the existence of the body-centered structure (α-Fe), the face-centered cubic structure (γ-Fe) and in very low concentration also of $Fe_3C$ could be verified. For a hyperthermia application a high yield of the ferromagnetic α-Fe is requested. By changing details of the synthesis process the relative yields of the different Fe-phases can be adjusted. The rate of the ferromagnetic α-Fe is increased by an annealing process below the temperature of the phase transition of γ-Fe to α-Fe which indeed yields a larger saturation magnetization [59]. To get exact information about the morphology of the iron filled CNT (γ-Fe/α-Fe ratio), the $^{57}$Fe Mössbauer spectroscopy is an appropriate method. Transmission Mössbauer spectroscopy (TMS) and backscattered conversion electron Mössbauer spectroscopy (CEMS) were applied in order to distinguish different Fe phases and their spatial distribution within the whole sample and along the tubes' height by the group of Ruskov. A characterization (on a large spatial scale) of the aligned CNT samples was performed by obtaining TMS spectra for selected spots positioned at different locations of the sample. While the total Fe content changes considerably from one location to another, the γ-Fe/α-Fe phase ratio is constant onto a relatively large area. Using TMS and CEMS for all aligned Fe-MWCNT samples it is also shown that along the CNT axes, going to the top of the nanotube the relative content of the γ-Fe phase increases. Going to the opposite direction, i.e., towards the silicon substrate, the relative content of the $Fe_3C$ phase increases [61].

4. Magnetic properties and hyperthermia feasibility

Various studies have shown that, independently of the synthesis technique, the carbon encapsulated iron is efficiently protected by the surrounding shells and its magnetic properties are retained. Often, magnetic characterisation studies have been performed for CNT fixed to a $SiO_2$ substrate. For such studies, Fe-CNT are grown directly on the substrate so that a rather dense coating of well aligned nanowires is studied. For such aligned Fe-filled CNT, uniaxial magnetic anisotropy is reported with the easy magnetic axis being parallel to the CNT axis. The anisotropy can straightforwardly attributed to the shape anisotropy of the magnetic nanowires. However, depending on the alignment [43] and density of CNT on the substrate magnetic dipole-dipole interactions between the Fe-cores have to be considered, too. Interestingly, enhanced magnetic coercive fields of $H_C$ >100 mT are observed in such ensembles of Fe-filled CNT at room temperature [37,43] which significantly exceed the value $H_C$ =9*10$^{-3}$ mT observed in bulk Fe. The more representative data for biomedical applications are achieved by powder measurements. Here, depending on the filling ratio, diameter, etc. coercive fields in the range of $H_C$ ~ 20-50 mT have been found. In a recent study the magnetisation of cells dried after incubation with Fe-filled CNT exhibited a critical field of similar size (~23 mT) [33].

Magnetic coercivity is even more reduced when the magnetisation is measured in a living cell culture [32]. For such a study, $2*10^5$ PC-3 cells were incubated with 50µg/ml of Fe-filled CNT. After an incubation period of 4h, the cells were washed two times with PBS, trypsinized and centrifuged at 4°C. This procedure in total yielded approximately $1*10^6$ cells which were resuspended in 100µl for media and measured, at 4°C, in a maximum applied field of H = 0.5T. Interestingly, the data imply only a vanishing coercivity of $H_C$ ~ (2±2) Oe.

The magnetisation data suggest the feasibility of Fe-filled CNT for by applying an alternating (AC) magnetic field (see e.g. [62]). The heating effect is based on the physical principle that applied AC magnetic fields induce magnetisation loops. If these loops are not completely reversible (e.g. in the case of ferromagnetic particles or for superparamagnetic particles below the blocking temperature), magnetic energy is transformed into heat. We mention that residual ferromagnetic catalyst particles appear in most synthesis processes which provide superparamagnetism in CNT (e.g.

[63]). Another route to superparamagnetic CNT employs the insertion of Gd ions to form Gadonanotubes [64].

There are various physical mechanisms which yield non-reversible loops, i.e. which are related to magnetic dissipation processes, and it should be emphasized that in the case of Fe-filled CNT the mechanism has not yet been identified. In general, different processes will be related to different energy scales and resonance frequencies, so that the filling material, its size and its shape have a significant effect on the heating properties. For example there will be strong differences between the case of single-domain particles, where the magnetisation process competes with shape and crystal anisotropies, and the motion of domain walls in multi-domain particles.

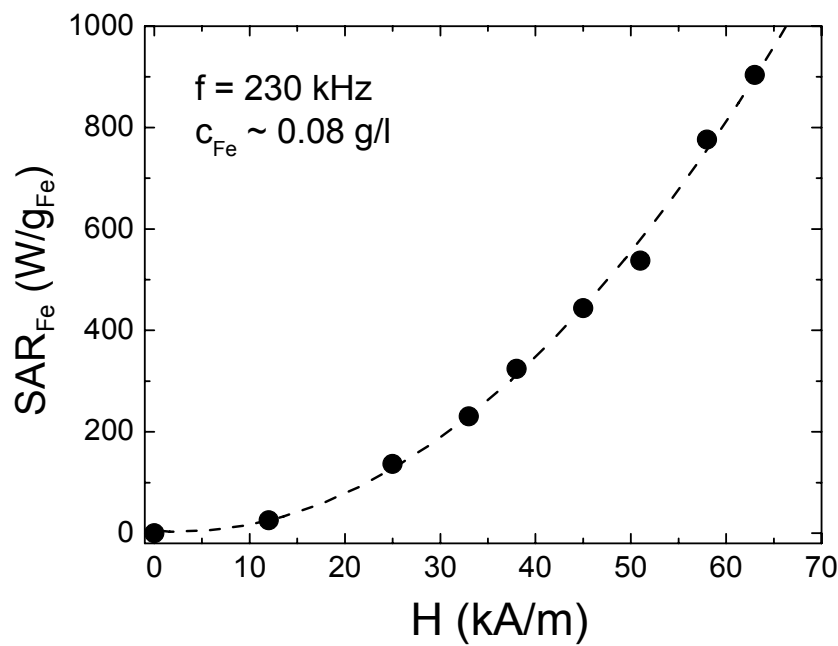

**Fig. 3**: Specific absorption rate $SAR_{Fe}$ of Fe-nanowires (~0.08 g Fe/l) encapsuled in CNT. Measurements have been performed at f = 230 kHz after dispersing the CNT by means of human albumin in PBS (4.2g/l).

The feasibility of Fe-filled CNT for hyperthermia has been demonstrated by a calorimetric study in AC magnetic fields. Our test device consists of a high-frequency generator with an impedance matching network and the magnetic coil system. Water-cooled copper tubes are wound into a coil system (e.g., 4–10 turns, diameter of bore 20–100mm) in which the sample is placed. Temperature is controlled by a fiber-optic temperature controller (LXT-Luxtron One). The generator can provide alternating magnetic fields in the frequency range of 50-1200 kHz. The data shown in Fig. 3 have been obtained at f = 230 kHz. Here, Fe-filled MWCNT have been dispersed by human albumin in PBS. The concentration was 4.2 mg/ml so that about 0.08 mg Fe/ml can be estimated. This concentration is corroborated by analysis of magnetization loops. The initial slope of temperature vs. time studies was used to determine the specific absorption rate SAR. In agreement with previous data [33], values of $SAR_{Fe}$ < 100 W/gFe were found in the field range below ~20 kA/m$^2$ which is applicable for medical treatment. We emphasize, however, that the presented first studies did not yet elucidate the actual dissipative mechanism. Detailed information about the magnetisation reversal will help to optimize e.g. the geometrical dimensions of the Fe-nanowires in order to achieve improved switching behaviour.

In addition to a magnetic field induced thermal ablation described in detail above, CNT are also feasible for near-infrad (NIR) light based hyperthermia. This method applies the fact that biological systems are transparent to light in the NIR regime of 0.7-1.1 $\mu$m wavelength while the strong absorbance renders single-walled CNT for hyperthermia agents in living cells [11,65,66]. Focused heat transduction and photo-ablative destruction of kidney cancer cells has also been shown for multi-walled CNT if being nitrogen doped [67]. Although NIR light is capable of passing through several centimetres of tissue, its interaction with biological matter is by far larger than of magnetic fields and the penetration depth is much smaller. For efficient heat transfer to deep tissue and in order to avoid parasitic heating effects in surrounding tissue magnetic agents hence seem to be advantageous.

## 5. Temperature control by NMR on filled CNT

Accurate control of the tissue temperature is mandatory in any hyperthermia approach. Currently, in the clinical treatments, temperature is controlled by a clinician's intervention by placing thermocouples or fiber-optical thermometers into the tumor in combination with computer modelling later on. Any model, however, demands estimates of the tissue properties, blood perfusion rate and other dynamic properties [68]. Instead, a continuous non-invasive temperature monitoring appears to be advantageous not only for hyperthermia treatment. Magnetic resonance (MR) thermometry based on a temperature-dependent proton resonance frequency shift of the water molecule provides temperature control in addition to a good spatial localization, thereby allowing for accurate target identification in ultrasound thermal therapy [69]. However, large doses of unshielded magnets which are present in nanoparticle-based hyperthermia introduce magnetic field inhomogeneities that reduce MRI contrastivity based on the proton relaxation-weighted image and prevent proton-based MR thermometry.

A promising approach for a non-invasive *in-vivo* temperature control relies on the use of a nanoscaled thermometer, which consists of a CNT and a filling material with strongly temperature dependent NMR parameters. In particular, the filling material might exhibit strong T-dependencies of the spin-lattice or the spin-spin relaxation, resonance frequency, dipolar or scalar couplings, and electrical quadrupole coupling at 310-350 K (ca. 20-60 °C) so that temperature detection is possible with high accuracy (<0.1 degree). Due to the protecting carbon shell, the number of materials which can be used for temperature sensoring without toxic adverse effects strongly increases. On the other hand, the container feature of CNT might, in principle, allow simultaneous filling of CNT with a temperature sensor and another probe such as a ferromagnet (=heater), thereby combining different functionalities in one kind of CNT (Fig. 1).

Up to know, the more simple system has been studied, where only the temperature sensor is encapsulated in CNT [53]. Many alkali and cuprous halides are known to show pronounced temperature dependencies of NMR parameters. From a family of these compounds monovalent cuprous iodide (CuI) turned out to be most suited. Here, both the copper and iodine nuclei have NMR active isotopes with a high natural abundance.

The synthesis of CuI filled CNT was based on pristine nanotubes consisting from 10 to 40 carbon layers with inner diameters between 5 and 20 nm. For filling with CuI, CNT were opened using thermal and acid treatment in combination with sonication. The opened CNT were grinded in a mortar gently and put in a silica glass ampoule together with CuI (Aldrich 99.99%) in excess. The ampoule was sealed under vacuum ($10^{-3}$ Torr) and heated at 600°C for 24h. At this temperature CuI is completely sublimated and transported into the opened CNT thanks to the capillary effect. The resultant material was examined by transmission electron microscopy (TEM and HRTEM), X-ray diffraction analysis and energy dispersive X-ray analysis (EDX) and identified as CNT filled to 80 % with single crystalline cuprous iodide.

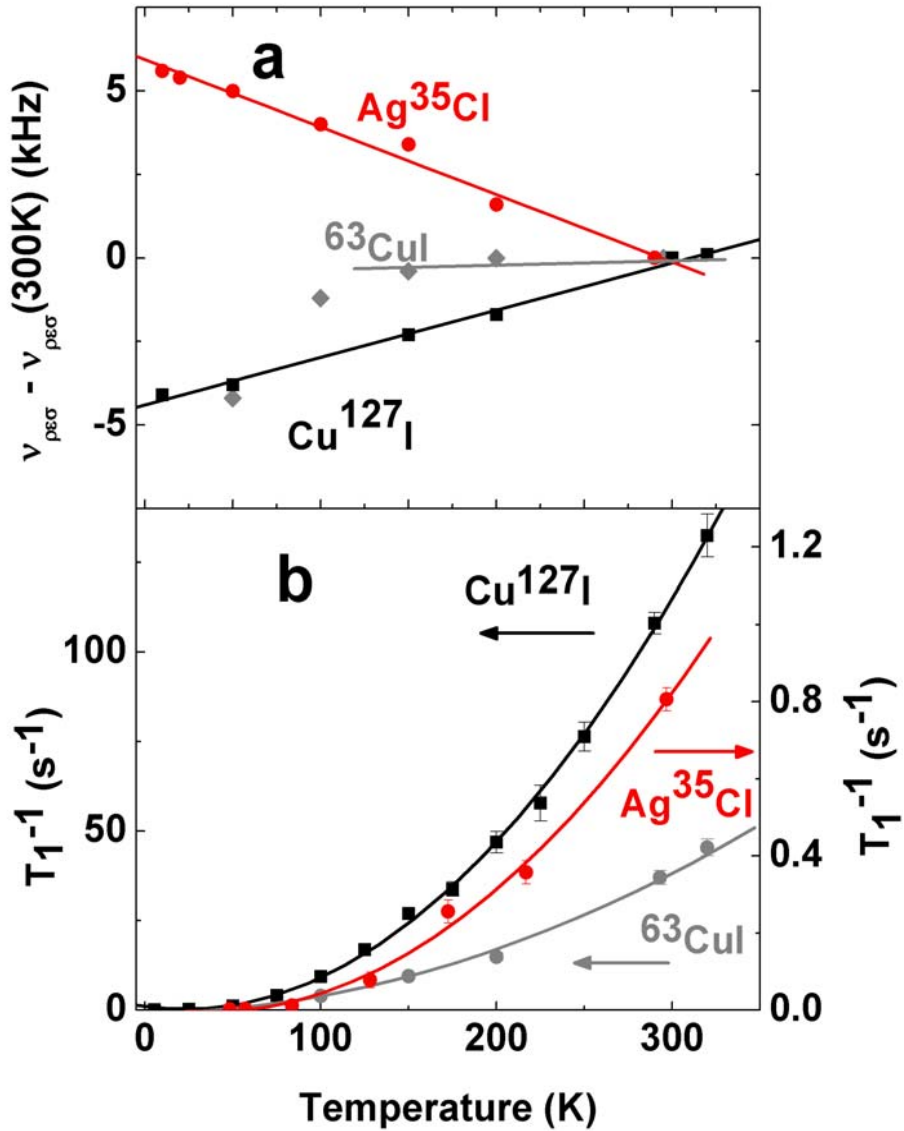

**Fig. 4:** Temperature dependencies of [127]I, [35]Cl and [63]Cu nuclear magnetic resonance parameters measured on CuI- and AgCl-filled CNT. (a) Nuclear magnetic resonance frequency. (b) Nuclear spin-lattice relaxation rate. The symbols present the experimental data. Solid lines are the fit (see the text).

Although both [63]Cu and [127]I nuclear isotopes possess a quadrupole moment, in CuI copper and iodine atoms each are surrounded tetrahedrally by four atoms of the opposite kind. This leads to a vanishing quadrupolar coupling. Therefore only the resonance frequencies, linewidths and relaxation times can be considered as temperature indicators. The NMR measurements were done in a standard solid state NMR spectrometer in the external magnetic field of 7.05 T. Both $^{63}$Cu and $^{127}$I NMR spectra obtained from the Fourier transformed echo signals represented a single resonance line. The spin-lattice relaxation $T_1$ was measured employing an inversion recovery pulse sequence. The obtained decay of $^{63}$Cu and $^{127}$I longitudinal magnetisation was analyzed following a standard equation for the spin I = 3/2 and I = 5/2, correspondingly. In both cases the magnetisation curves followed a simple exponential form characterized by a single spin-lattice relaxation time.

The analysis of the $^{63}$Cu NMR spectra reveals insignificant changes in the resonance frequency in the relevant temperature range, while the resonance frequencies measured on the $^{127}$I nucleus indicate a stronger dependence on temperature (Fig. 4a). Such behaviour is explained by lattice effects which include both the lattice vibration and the lattice expansion. The $^{127}$I NMR frequency data are well fitted with a linear function over the whole temperature range studied. Thus, at first glance, the $^{127}$I resonance frequency might be used as a measure for the temperature determination. A relative small slope of this function leads, however, to an error of 15 K in temperature determination. Therefore the usage of this parameter for an accurate temperature control is not reasonable. Furthermore, the spectral linewidths at half maximum have been analyzed for both nuclei and were found to be constant at temperatures below 320 K. Thus, also the linewidths could be ruled out to serve as a temperature control parameter.

The temperature dependence of the $^{127}$I spin-lattice relaxation rates $T_1^{-1}$ is presented in Fig. 4b. The $^{63}$Cu $T_1^{-1}$ dependence is similar but much smaller compared to the $^{127}$I data. The $T_1^{-1}$ dependencies for both nuclei are found to be in very good agreement with the law $T_1^{-1} \sim T^2$ that is expected for a Raman two-phonon quadrupolar process. [70]. This behaviour is observed over the entire temperature range implying no contributions from impurities which might appear at low temperatures and from ionic diffusion which might be observed at high temperatures. This is consistent with the view that relaxation is driven by a quadrupolar mechanism in this compound [71]. The $^{127}$I experimental data are well fitted with a quadratic function $T_1^{-1} = a + bT + cT^2$, where fitting coefficients are a = 1, b = $(7\pm1)*10^{-2}$ and c = $(1.49\pm0.05)*10^{-3}$. The mean squared errors of the fitting coefficients provide an estimate of the accuracy of the temperature determination. In the temperature range of biological interest (i.e. 290 to 320 K) the CuI-CNT nanothermometer can indicate the temperature with an accuracy of 2 K by means of the spin-lattice relaxation measurement. Other Cu-halides filled in CNT demonstrate a qualitatively similar behaviour but exhibit less temperature sensitivity as shown in table 1. These results, in particular those on CuI-CNT, provide a good starting point to look for further filling materials of CNT in order to increase the accuracy of temperature determination.

| Material | Nucleus | d$\nu_{res}$/dT (Hz/K) | d($T_1^{-1}$)/dT (Hz/K) |
|---|---|---|---|
| CuI-CNT | $^{63}$Cu | - | 0.27 |
| | $^{127}$I | 14 | 0.86 |
| CuBr-CNT | $^{63}$Cu | 19 | 0.15 |
| | $^{81}$Br | 7 | 0.75 |
| CuCl$_2$-CNT | $^{63}$Cu | 15 | 0.23 |
| | $^{35}$Cl | 1 | - |
| AgCl-CNT | $^{35}$Cl | 21 | 0.006 |

**Table 1:** Temperature sensitivity parameters of several filled CNT. The table shows the filling material, the respective nucleus as well as temperature dependence of resonance frequency and $T_1$-relaxation time in the temperature range of 300-320 K.

Conclusions

Due to their extraordinary physical and chemical properties carbon nanotubes reveal a promising potential for applications on the cellular level. Upon filling, nanoscaled containers are realised in which the active material is encapsulated by a protecting carbon shell. We describe different synthesis routes and show the structural and magnetic properties of CNT. In particular, the filling with magnetic material offers the potential for hyperthermia applications while the insertion of NMR active substances allows the usage as markers and sensors. This potential of carbon nanotubes for biomedical applications is highlighted by hyperthermia studies which prove their applicability for local in-situ heating. In addition we have shown that a non-invasive temperature control by virtue of a carbon-wrapped nanoscaled thermometer is possible.

A valuable extension would be spatially resolved NMR. The rapidly growing field of cellular and molecular MR imaging enables to visualize cells and inserted CNT in order to track cancer cells and to control therapies on the cellular level such as magnetic hyperthermia. The container feature of CNT is extensively utilized if a heating element (ferromagnet), a temperature sensor and a contrast agent are confined within the same nanocontainer. Such a combination of different functionalities on a nanoscale would provide simultaneous heating, temperature control by means of MRI and high spatial resolution of the image. Furthermore, if clinical usage of the static magnetic field is contra-indicated and MR imaging is no longer suitable then the technology proposed here addresses materials with temperature dependent nuclear quadrupolar resonance NQR (e.g. cuprous oxide) or zero-field NMR (e.g. Co-based compounds) parameters that demonstrates versatility of this approach for biomedical applications. The potential of CNT for biomedical applications becomes even more evident if their container function is exploited for drug delivery in magnetically functionalised and NMR labelled nanodevices. It has been shown that anti-cancer drugs can be inserted in CNT [54] for which purpose the multi-walled ones turn out to be particularly appropriate. Combing different functionalities in well shielded containers hence seems to be the particular advantage of carbon nanotubes.

Acknowledgements

This work was partly supported by the European Community through the Marie Curie Research Training Network CARBIO under contract No. MRTN-CT-2006-035616. The authors thank Diana Haase, Manfred Ritschel, Anastasia Vyalikh, Anja Wolter, Kamil Lipert, Yulia Krupskaya, Christopher Mahn, Thomas Mühl, Dieter Elefant and Hans-Jörg Uhlemann from IFW Dresden as well as Arthur Taylor and Kai Krämer from the Department of Urology, Technical University Dresden. We particularly appreciate Albrecht Leonhardt for support and valuable discussions. Scientific advice by Sabine Achten (Boehringer Ingelheimer Fonds), Francois Rossi (European

Commission Joint Research Centre, Ispra) and Andreas Jordan (Magforce Nanotechnologies AG) is gratefully acknowledged.